\documentclass[twocolumn,preprintnumbers,amsmath,amssymb]{revtex4}

\usepackage{graphicx}
\usepackage{dcolumn}
\usepackage{bm}

\begin{document}


\title{Close packing density of polydisperse hard spheres}

\author{Robert S. Farr}
 \email{robert.farr@unilever.com}
 \affiliation{Unilever R\&D, Olivier van Noortlaan 120, AT3133, Vlaardingen, The Netherlands}

\author{Robert D. Groot}
 \email{rob.groot@unilever.com}
 \affiliation{Unilever R\&D, Olivier van Noortlaan 120, AT3133, Vlaardingen, The Netherlands}

\date{\today}

\begin{abstract}
The most efficient way to pack equally sized spheres isotropically in 3D is known as 
the random close packed state, which provides a starting point for many 
approximations in physics and engineering. However, the particle size distribution of 
a real granular material is never monodisperse. Here we present a simple but accurate 
approximation for the random close packing density of hard spheres of any size 
distribution, based upon a mapping onto a one-dimensional problem. To test this 
theory we performed extensive simulations for mixtures of elastic spheres with 
hydrodynamic friction. The simulations show a general (but weak) dependence of 
the final (essentially hard sphere) packing density on fluid viscosity and on particle 
size, but this can be eliminated by choosing a specific relation between mass and 
particle size, making the random close packed volume fraction well-defined. Our 
theory agrees well with the simulations for bidisperse, tridisperse and log-normal 
distributions, and correctly reproduces the exact limits for large size ratios. 
     
\end{abstract}

\pacs{XXX}

\maketitle

\section{Introduction}     
Granular materials such as sediments and powders are widespread in nature and 
industrial contexts, and treating the grains as hard spheres is often a useful first 
approximation. In these systems, the manner in which the grains pack together has 
profound influence on properties. Of these, random close packing\cite{1} is most likely to be 
encountered in tapped and consolidated systems, although other possibilities, such as 
random loose packing\cite{2}, and various crystalline arrangements (whose existence is very 
sensitive to the form of the size distribution and method of creation\cite{3,4}) are also possible.

Even though the precise nature (and for monodisperse spheres, even the existence\cite{5}) 
of the random close packed state remains the subject of ongoing research\cite{6}, it provides a 
starting point for many approximations in both physics\cite{7,8} and engineering, and has great 
practical importance not only for the prediction of the density of granular materials, but 
also other properties. For example, the viscosity of dense dispersions will diverge at this 
point\cite{9} and it is related to the permeability in packed beds\cite{10}.

One of the insights that have come forward from simulations\cite{6}
is that the dense random packing density of hard spheres depends upon the (shear) friction
coefficient, if the particles only lose energy by inelastic collisions. Further, the 
jamming density of a hard sphere system depends upon the initial state, and on the particular pathway 
chosen to cool down the system. In general, however, dissipative interactions play a role not
just at contact. Granular particles suspended in a viscous medium also dissipate energy via 
long-range hydrodynamic interactions. Hence we anticipate that the dense random packing also depends
upon solvent viscosity and on the range of the (hydrodynamic) friction. This is the first problem 
that we wish to address. To this end we developed a new
simulation method that includes these effects. Using this method we not only find that the dense 
random packing depends on fluid viscosity, but -- quite unexpectedly -- also on particle size and mass. 
By analysing the various time scales in the problem we obtain a way to eliminate this dependence, 
which sheds new light upon the nature of the dense random packed state.
 
\begin{figure}
\includegraphics[width=2.5in]{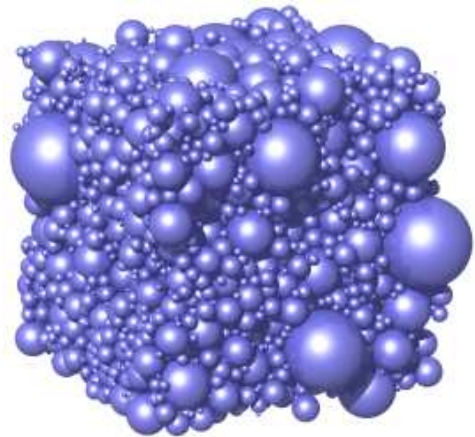}
\caption{\label{fig1}
Close packed configuration of spheres from a log-normal 
distribution. The spread in the logarithm of radius is $\sigma = 0.6$. Only spheres with centres lying 
in one periodic image of the simulation cell are shown. 
}
\end{figure}

In practical cases the particle size distribution of a real granular material, or 
mixture of materials, is never monodisperse. Also for such polydisperse problems modelling 
techniques\cite{11,12} have been used to calculate maximum packing fractions of spheres 
with a distribution of sizes. A typical example of a polydisperse system in a close packed 
state is shown in Fig. \ref{fig1}. 

The second problem that we wish to address is that these simulation methods are quite time 
consuming, and therefore their applicability is limited. It would be desirable to have an 
analytic expression for the close packing density, or a fast approximation, but progress in 
this direction has not been rapid. Ouchiyama and Tanaka\cite{13} have presented a theory based 
upon the volume occupied by a sphere in contact with other spheres of the mean diameter, 
but their results are at best qualitative, and the reasoning behind the method is not simple 
enough to suggest obvious improvements. Song {\em et al}.\cite{6} presented a theory for the packing of 
monodisperse spheres, but the generalisation to an arbitrary size distribution is not obvious. 
Recently Biazzo {\em et al}.\cite{14} presented a theory for binary mixtures, 
but like the Ouchiyama-Tanaka theory, it violates the exact upper limit 
for large fractions of big spheres that is 
given in Eq. (\ref{eq3}) below. Thus, a comprehensive theory to predict the random close packing 
density of an arbitrary sphere mixture is still lacking. 

We formulate such a theory, which is presented here in Section II. Next, we define our 
simulation method in Section III, and 
we present the combined theoretical and simulation results in Section IV. Conclusions are 
formulated in Section V.

\section{Theory}
Here we propose an approximate solution to the problem of polydisperse packing density, 
obtained by abstracting what we believe to 
be essential features of the physics and geometry of packing. The fundamental problem we 
wish to solve is as follows: suppose we have a normalised distribution $P_{3D}(D)$ of sphere 
diameters defined so that $P_{3D}(D){\rm d}D$ equals the number fraction of spheres with diameters in 
the range $(D, D+{\rm d}D)$ present in the system. Then we ask what is the 
functional ${\cal F}:P_{3D}(D)\mapsto\phi_{\max}$
that maps the size distribution onto the random close packed volume 
fraction? 

\begin{figure}
\includegraphics[width=3.0in]{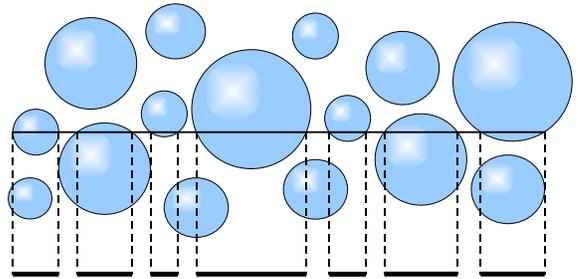}
\caption{\label{fig2}
How to map a 3D sphere distribution onto a rod distribution. A 
straight line through a random arrangement of spheres defines a set of rods. 
The probability that a sphere is hit by the line is proportional to its 
cross sectional area and so proportional to $D^{2}P_{3D}(D)$. 
The probability that such a hit produces a rod of 
length $L$ is $2L/D^{2}$ (for $L<D$), and Eq. (\ref{eq1}) in the text follows.
}
\end{figure}

In order to construct this functional, we begin by mapping the 3D 
sphere packing 
problem onto a packing problem of rods in 1D. The corresponding 1D distribution 
$P_{1D}(L){\rm d}L$ gives the number fraction of rods with rod length in the range $(L, L+dL)$ present 
in the system. To do this we imagine a large random, but non-overlapping, arrangement of 
spheres in 3D with size distribution $P_{3D}(D)$. This need not be close packed for the argument 
that follows. Now imagine drawing a straight line through this distribution, and counting 
each portion of the line which lies within a sphere as a rod (see Fig. \ref{fig2}). The resulting 
distribution of rod lengths is then given by
\begin{equation}\label{eq1}
P_{1D}(L)=2L\frac{\int_{L}^{\infty}P_{3D}(D){\rm d}D}{
\int_{0}^{\infty}P_{3D}(D)D^{2}{\rm d}D}.
\end{equation}

If rods of length $L_{i}$ are arranged on a line of length $\Lambda$ 
(with periodic boundary conditions), 
then the rod length fraction is clearly given by $\psi=\Lambda^{-1}\sum L_{i}$. 
This equals the volume fraction $\phi$ if there is a corresponding system 
of spheres in 3D.
   
Let us assume that it is possible to map the closest random packing of spheres in 3D 
onto a problem of packing the above collection of rods on a line, where we search over all 
orderings of rods as well as their separations. This mapping will be achieved through an 
effective potential between the rods, which must have the following properties: 

1. It should lead to a maximum packing fraction $\psi_{max}$ which is unchanged if all the rods (or 
spheres) are magnified by an equal amount; 

2. The potential should be `hard', in that it is either zero or infinite;

3. The interaction potential between large rods should reach through small rods. This will 
allow very small rods to `rattle around' in the gaps between the large rods, so that at high 
weight fractions of large rods the latter can form a stress-bearing network;

4. The interaction range for mixtures of very unequal rods should be determined by the size 
of the smallest rod, so that small rods can form a dense randomly packed system in between 
the large rods, without leaving large gaps.

The true interaction potential between the rods will be both many-body and highly 
complicated, capturing topological aspects of 3D space. However, we suggest that the 
following pair potential is the simplest expression which honours the four requirements 
listed: 

If two rods $i$ and $j$ have a gap $h$ between their nearest approaching ends, then we 
introduce the potential
\begin{equation}\label{eq2}
V(h)=\left\{
\begin{array}{lll}
\infty & {\rm if} & h < \min(fL_{i},fL_{j}) \\
0 & {\rm if} & h \ge \min(fL_{i},fL_{j}) 
\end{array}\right.
\end{equation}
where $f>0$ is a `free volume' parameter. If this potential applies between any pair of rods, 
regardless of intervening rods in the gap between them, then it will satisfy requirement 3. 
The other requirements follow naturally. Rather than taking the minimum of the two 
lengths one could introduce a more complicated function of $L_{i}$ and $L_{j}$, but the present choice 
appears to be the simplest to capture the physics. In the remainder of this article $f$ will be used
as a fit parameter; it is the only parameter in our theory, and can be chosen by requiring that
the theory reproduces the random close packing of monodisperse systems.

For each ordering of the rods on the line, there is a shortest line which can 
accommodate them without incurring an infinite potential energy, and this leads to a close 
packing fraction for this ordering, which is simply $\psi=\Lambda^{-1}\sum L_{i}$. The maximum packing 
fraction $\psi_{\max}$ is the maximum value attained by $\psi$ over all possible orderings of the 
rods. If all the $L$'s are equal, $\psi_{\max} = (1+f)^{-1}$, and any rod polydispersity (which will
always be present if we use Eq. (\ref{eq1})) will increase $\psi_{\max}$. 

If we imagine inserting rods one at a time to form a packing, while increasing $\Lambda$ if 
necessary, then Eq. (\ref{eq2}) constitutes a two-body potential between the inserted rod and 
all the rods currently in the packing. However, if rods are inserted in decreasing order of 
size, then the special choice of this potential means it only depends upon the newly inserted 
rod, and the packing further away is not disrupted by this process. To insert the new rod 
with a minimum increase of line length $\Lambda$ we need to identify the biggest gap. Therefore the 
following `greedy algorithm'\cite{15} may be used to find $\psi_{\max}$ for arbitrary $\{ L_{i}\}$:

(a) The set of lengths $\{ L_{i}\}$ is labelled such that 
$L_{1} \ge L_{2} \ge \ldots \ge L_{N}$. These will be inserted in 
decreasing order of lengths into the growing optimal packing, 
starting with $L_{1}$.

(b) Throughout the algorithm, we maintain a set of gaps $\{ g_{i}\}$, equal in number to the 
number of rods we have inserted into the packing. At the start, when we have only one rod, 
this set contains one element $g_{1} = fL_{1}$.

(c) In order to insert rod $j$, we identify $g_{\max}$, the largest gap in the set of gaps, and we 
remove it from the set. We then add two new gaps to the set, namely $fL_{j}$ and $\max[g_{\max} - (1+f)L_{j}, fL_{j}]$. 

This process implicitly increases the line length $\Lambda$, if that is 
needed to accommodate the new rod. 

Our final approximation consists of choosing a large number of rods from the 
distribution $P_{1D}(L)$, and packing them by the greedy 1D packing algorithm to obtain our 
estimate $\psi_{\max}$ for $\phi_{\max}$, the close packed volume fraction of the sphere distribution $P_{3D}$. 
Since this is essentially a sorting routine, the predictions of this 
algorithm take much less time (ca 0.3 seconds on a modern desktop computer) 
than explicit 3D simulation (1 to 30 hours). When the rod lengths are chosen 
at equidistant values of the cumulative 1D distribution, 2000 rods are sufficient for 
5 decimal places accuracy. Thus, for $N$ rods we choose rod length $L_i$ such that
$\int_{L_i}^{\infty}P_{1D}(l){\rm d}l = (N-i+1/2)/N$. We use $N=20000$ rods.

One useful property of this procedure is that it correctly reproduces the exact solution 
for bidisperse spheres with infinite size ratio. This limit is given by
\begin{equation}\label{eq3}
\psi_{\max}=\min\left(
\frac{\phi_{RCP}}{1-w(1-\phi_{RCP})},\frac{\phi_{RCP}}{w}\right),
\end{equation}
where $\phi_{RCP}$ is the maximum packing fraction for a {\em monodisperse} 
system, and $w$ is the mass fraction of large spheres on the total particle volume, so 
$w = \phi_{\rm large}/(\phi_{\rm large}+\phi_{\rm small})$. Numerical solution of the theory 
for size ratios down to 1:1000 shows minor ($\sim 1\%$) 
deviations from the exact limit, for $w$ values very close to the cusp, $w=1/(2-\phi_{RCP})$.

The description above is complete, except that we need to specify the parameter $f$, 
which should be chosen such that the predicted maximum packing for monodisperse
spheres is the correct random close packing value $\phi_{RCP}$. 
For monodisperse spheres we have $P_{1D}(L) = 2LD_{0}^{-2}\theta(D_{0}-L)$
(where $\theta$ is the Heaviside step function), and we find that a value 
of $f = 0.7654$ leads to a packing fraction of approximately $0.6435$. This 
value agrees with 
the simulation result described below, and is our only fit parameter.

\section{Simulation}
Several methods have been proposed in the literature to simulate the dense random packing of hard
spheres. One method used frequently was introduced by Lubachevsky and Stillinger,\cite{21} who
simulated hard spheres by Molecular Dynamics and slowly compress the system until it jams.
There is a drawback to this method, namely that ever smaller time steps need to be taken as close
packing is approached. Moreover, the physical relaxation time of the system diverges near the close 
packing density,\cite{26} and consequently long runs are necessary. 
To circumvent this problem O'Hern {\em et al.}\cite{16} used soft spheres, and 
located the minimum energy by a conjugate gradient (CG) algorithm. This method 
is much faster than a hard sphere simulation, but the disadvantage is that the CG algorithm 
only simulates the high friction limit.

In general the dense random packing dependends on friction,\cite{6} and  
dissipative interactions may play a role not
just at hard sphere contact. Granular particles suspended in a viscous medium also dissipate energy via 
long-range hydrodynamic interactions. Hence we anticipate that the dense random packing also depends
upon solvent viscosity and on the range of friction. Therefore we introduce a new
simulation method to include these effects. 

Following O'Hern {\em et al.}\cite{16} and Groot and Stoyanov\cite{17} we simulate repulsive elastic 
spheres in the limit $T \rightarrow 0$. 
The generalization of the repulsive force in this model to elastic spheres of unequal size is 
\begin{equation}\label{eq4}
F_{ij}^{Rep}=2ER_{ij}(D_{ij}-r)\theta(D_{ij}-r).
\end{equation}
Here, $r$ is the distance between particle centres, $D_{ij} = (D_{i}+D_{j})/2$ 
is the mean diameter and $R_{ij}$ 
is the harmonic mean radius given by $R_{ij} = \frac{1}{2}D_{i}D_{j}/D_{ij}$. 
The parameter $E$ is proportional to the linear elastic modulus of the particles\cite{17}. 
Henceforth we use $E = 1000$.

Instead of using a CG algorithm to search the energy minimum, or imposing energy dissipation at
particle collision, 
we introduce a soft friction function of finite range that represents the hydrodynamic 
interaction between spheres. A soft friction has been introduced before to simulate hydrodynamics in 
fluids, in the context of Dissipative Particle Dynamics\cite{27,19,20}. 
However, application to particles of unequal size 
is new to our knowledge, and because the dense random packing depends on friction some care must be
taken in defining the friction function.

The most general distance-dependent friction is
\begin{equation}\label{fric1}
{\rm\bf F}_{ij}^{frict}=-\gamma_{ij}{\rm\bf v}_{ij}^{r}g(r/r_c)
\end{equation}
where $\gamma_{ij}$ is a friction factor that may depend on both particle sizes, 
${\rm\bf v}_{ij}^{r}$ is the radial velocity difference, $g(r/r_c)$ 
is a distance dependent function, and $r_c$ is a cut-off distance that may again depend on particle
size. 

\begin{figure}
\includegraphics[width=2.5in,angle=90]{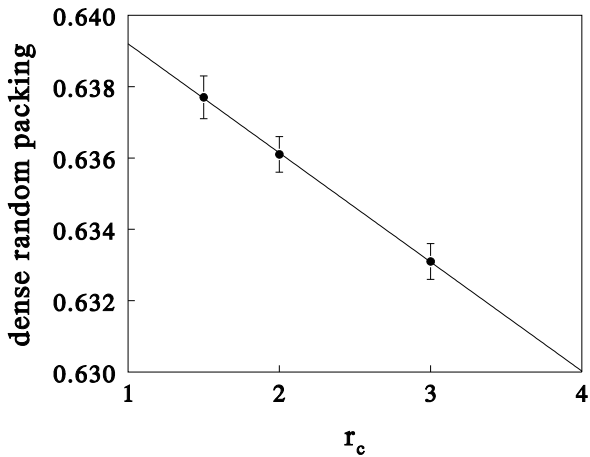}
\caption{\label{newfig3}
Volume fraction obtained by extrapolating the $P-\phi$ curve to zero pressure.
}
\end{figure}

To demonstrate the importance of the friction function, we first concentrate 
on monodisperse systems and study the influence of friction range and strength, 
and of particle size. We use a periodic $10\times 10\times 10$ box, containing from
1222 up to 1290 particles. All particles have diameter 1 and mass 1, and interact with a 
repulsive force $F = 10^3(1-r)$ for $r < 1$. 
First we study the influence of the range of the friction interaction. To this end we use the 
friction interaction 
${\rm\bf F}_{ij}^{frict} = -{\rm\bf v}_{ij}^r (1-r/r_c)^2/(1-1/r_c)^2 $; 
and the range is varied as $r_c =$ 1.5, 2, 3. With 
this choice for the friction force, the friction at particle contact is unity for all systems. 
All systems are evolved over $10^4$ steps or more, with $\delta t = 0.01$. For 
each system the pressure at T = 0 was averaged over 5 independent starting configurations.

Even though all systems have the same friction strength at $r = 1$, the mere range of friction  
appears to influence the pressure in the final state. As the friction range 
increases, so does the pressure at T = 0. In particular, the volume fraction to which the 
pressure extrapolates to zero -- the dense random packing -- varies systematically with the 
force range, see Fig. \ref{newfig3}. Although the effect is not very large (about 1\% variation), it is 
clear that the friction range does have an influence. The extrapolated value to $r_c = D = 1$, 
$\phi_{cp} = 0.6392\pm 0.0004$, compares well with the reported mean-field result for hard spheres 
with a friction interaction at contact,\cite{6} $\phi_{RCP} = 0.634$. 
Thus we conclude that, to eliminate the influence of the friction range, the range must be scaled
relative to the particle size. 

Next, to demonstrate the influence of particle size and friction strength, a series of 
simulations was done where the ratio of the friction range relative to the particle diameter 
was kept fixed at $r_c/D$ = 1.5. The friction force used in this study was 
${\rm\bf F}_{ij}^{frict} = -\gamma_f {\rm\bf v}_{ij}^r (1-r/r_c)^2 $, 
where $\gamma_f$ is a fixed friction factor, to be specified as input variable. 
Two sizes were studied, $D = 1/2$ and $D = 1$, and two friction factors were used, 
$\gamma_f = 1$ and $\gamma_f = 4$. In these simulations all 
particle masses were put at $m = 1$. The box size was taken as $V = 8^3$, and the conservative 
force was taken as $F = 10^3D(D-r)$ for $r<D$, the same as in the previous simulations. 

\begin{figure}
\includegraphics[width=2.4in,angle=90]{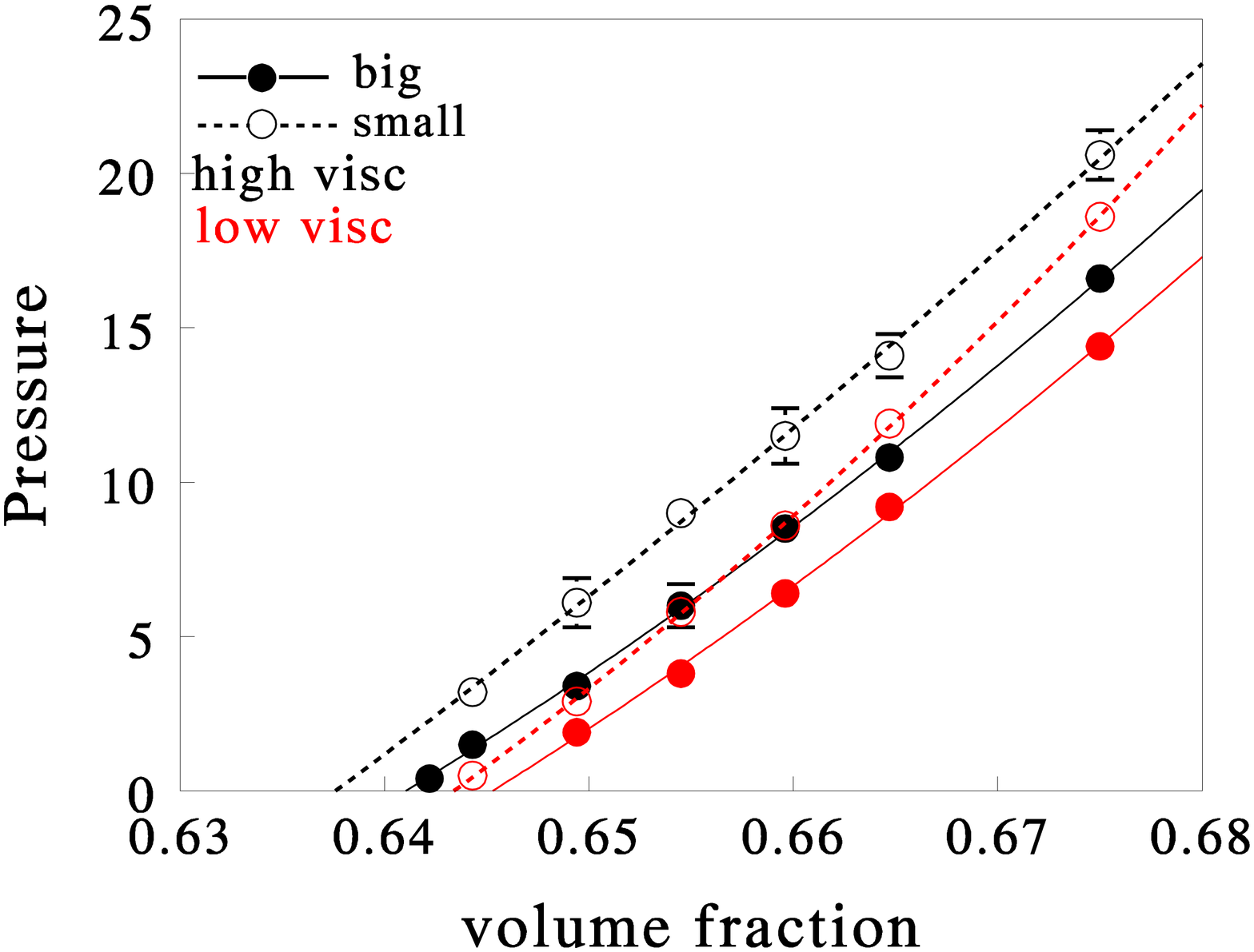}
\includegraphics[width=2.4in,angle=90]{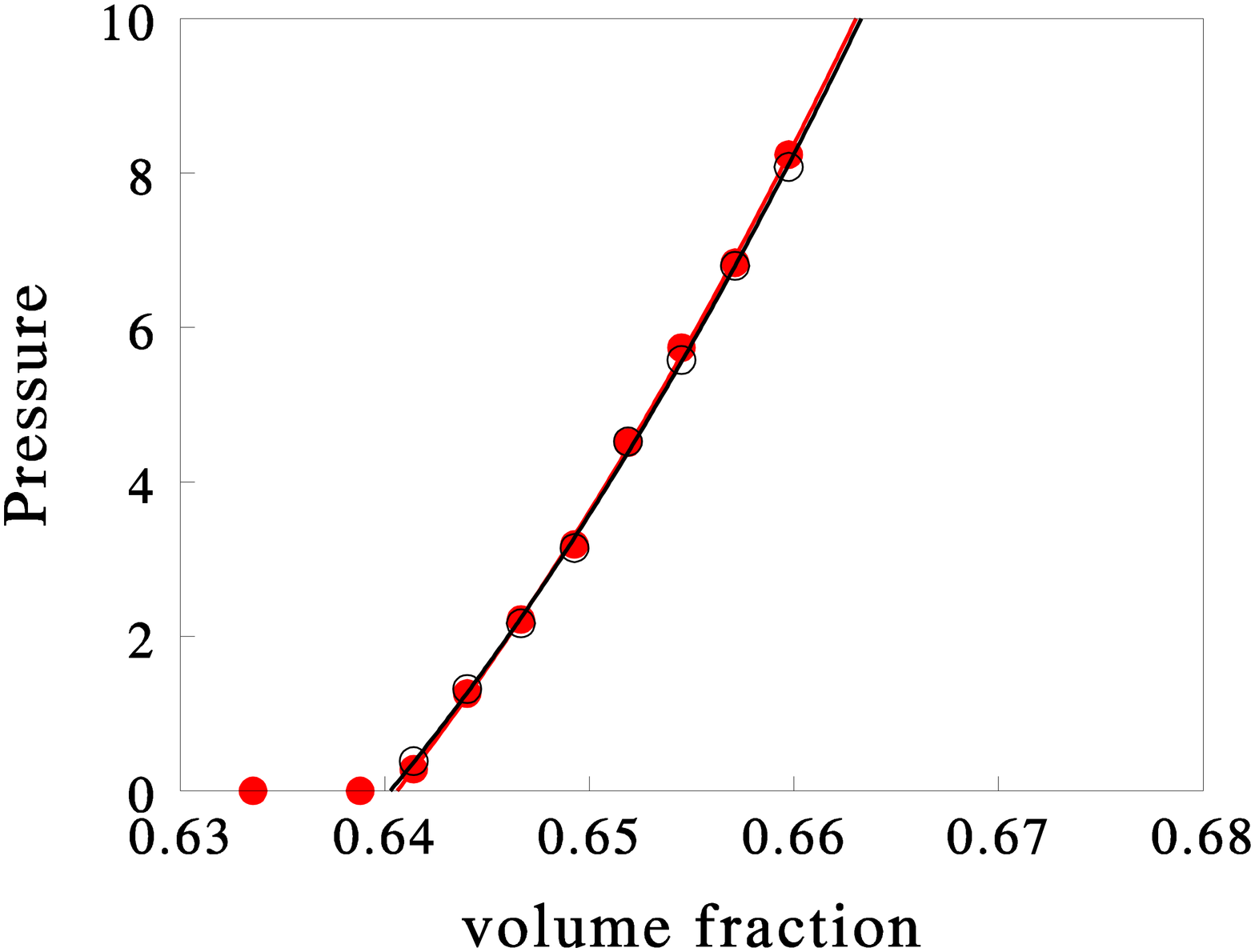}
\caption{\label{newfig4}
Mean pressure at T = 0 as function of particle volume 
fraction; a (top), for big particles (full symbols) and small particles (open symbols) and for
low (red) and high (black) viscosity;  b (bottom) big particles (red dots) and small particles 
(black circles) using friction as in Eq. (\ref{eq5}) and mass $m=D$. }
\end{figure}

The results are shown in Fig. \ref{newfig4}a. The red lines give the results of low 
viscosity ($\gamma_f = 1$) and 
the black lines give the results of higher viscosity ($\gamma_f = 4$). 
Results for small particles are denoted by 
open symbols and dashed curves, while results for big particles are denoted by closed 
symbols and full curves. This shows that both the friction factor and particle size influence the 
pressure in the glassy state. Consequently, the dense random packing density must depend on 
particle size. Even though the effect is small it is important, as it points at a reason why the
dense random packing density is ill-defined. For practical reasons we wish to define
a dense random packing density that does not depend on particle size, i.e. that is scale invariant.
To obtain this, it is not sufficient to have a friction range that is proportional to the 
particle diameter; the packing density depends in a complicated way on particle size and on the
strength and range of the friction force. Empirically there may 
seem to be some scaling when the friction is increased with the square of particle size (results
for $D = 1/2$, $\gamma_f = 1$ in Fig. \ref{newfig4}a partially overlap with $D = 1$, $\gamma_f = 4$), 
but the slopes of the curves are clearly different. 
 
The above results show that we cannot just take any friction function. It
must reflect the physical properties of the hydrodynamic interaction. One 
physical property is the scaling of 
the hydrodynamic force with particle size. On dimensional grounds the friction 
factor must be proportional 
to particle radius, as in Stokes' law, $F=6\pi\eta av$. More generally, the (radial)
squeeze mode of the hydrodynamic force 
between two rigid spheres {\em at close contact} behaves as\cite{18} 
\begin{equation}\label{neweq5}
{\rm\bf F}_{ij}^{frict} = -\gamma {\rm\bf v}_{ij}^{r} R_{ij} g(h/R_{ij}) 
\approx -\gamma {\rm\bf v}_{ij}^{r} \frac {R_{ij}^2} h
\end{equation}
where $\gamma$ is the friction factor that is proportional to fluid viscosity, and $h=r-D_{ij}$ is the 
distance of closest approach. The function $g(x)\sim(1/x)$ is a scaling 
function that represents the lubrication force.

There is a large body of evidence showing that correct long-range inertial 
hydrodynamics is generated even if the (divergent) lubrication force between particles is replaced 
by a finite distance-dependent friction\cite{19,20}. 
It is important however to choose the harmonic mean radius as the scaling length (unlike the 
conjecture by Kim and Karilla\cite{18}), otherwise the friction between very unequal 
spheres would vanish if we remove the divergence of $g(x)$. Thus, we use the friction function
\begin{equation}\label{eq5}
{\rm\bf F}_{ij}^{frict}=-\gamma{\rm\bf v}_{ij}^{r}R_{ij}
(1-h/R_{ij})^{2}\theta(R_{ij}-h).
\end{equation}
which captures the right physics regarding the scaling of the range and strength of the viscous 
interaction with particle size.
 
Now we can turn to the problem of defining a size invariant dense random packing density.
Therefore we 
analyse the relevant time scales of the problem. For a monodisperse system of elastic 
particles, a first time scale is the oscillation time, 
$t_{el} = 2\pi(m/ED)^{1/2}$. This is the elastic time 
scale. The second time scale in the system is the drag relaxation time, 
$t_{d} = m/(\gamma D)$. The dense 
random packing can only be independent of particle size if we maintain a 
constant ratio between these two time scales, 
$t_{el}/t_{d} = 2\pi\gamma(D/Em)^{1/2}$. Thus, we have scale invariance only 
if the friction factor satisfies $\gamma\propto (m/D)^{1/2}$. Since for most 
systems mass scales as $m \propto D^{3}$, this implies that for (soft) 
elastic spheres with hydrodynamic interaction the dense random packing 
fraction is (weakly) particle size dependent. 

To obtain a well-defined dense random packing we are forced to choose the particle mass $m$
proportional to $D$. When this choice is made, the above ratio of time 
scales becomes particle size 
independent, and consequently the dense random packing fraction is well-defined. This 
choice has been made henceforth. The predicted scaling was checked by simulation and 
holds exactly. The pressure as function of time for systems of particle diameter $D = 1/2$ and 
$D = 1$ fall on top of each other if we scale the friction range with particle size 
(as in Eq. (\ref{eq5}))
and simultaneously impose $m\propto D$. To demonstrate the improvement in system definition,
two series of simulations were done, again for monodisperse systems 
of particle diameter $D = 1/2$ and $D = 1$, with repulsive force $F = 10^3 D(D-r)$ for $r < D$. For a 
realistic hydrodynamic scaling we used Eq. (\ref{eq5}), with friction factor $\gamma = 0.74$. 
Because the systems are 
monodisperse the cut-off distance for the friction interaction is $r_c = 1.5D$, as in the previous 
case. To obtain scale invariance, we choose the masses as $m = D$. 
The systems had fixed volume $V = 10^3$ for 
$D = 1$ and $V = 5^3$ for $D = 1/2$. The systems were integrated over 5000 steps with time step 
$\delta t = 0.01$ and the pressure was averaged over 10 independent runs. The comparison between 
small and big particles is shown in Fig. \ref{newfig4}b. The predicted scaling is followed excellently.

Now that the system is well-defined, we can define a fast algorithm to obtain 
the close packing density. We use a variation of the Lubachevsky-Stillinger 
algorithm\cite{21}, where we make use of the relative 
softness of the interaction potential. We prepare the system in a random conformation 
(with particle overlaps) and 
then evolve it in an $(N, V, T)$ ensemble until we have a completely equilibrated state. For the 
parameters $\gamma = 1$ and $\delta t = 0.01$ that we used, this requires 
$3-50\times 10^3$ time steps. Then we switch to an $(N, P, T)$ 
ensemble, where the pressure is steered towards $P = 0.01$, which is close 
enough in practice to $P = 0$ (the error in $\phi_{\max}$ is of the order 
$10^{-5}$). If during a run the pressure falls below $0.001$ we switch to the L-S algorithm 
and compress the system in small steps until 
the pressure turns positive. The advantage of this method over the standard L-S algorithm is that the 
(high) pressure in the initial $(N, V, T)$ simulation quickly drives the system towards $P = 0$. 
The final $(N, P, T)$
simulation serves to run down the $P-\phi$ curve (see Fig. \ref{newfig4}b) to locate the intercept at
$P=0$. Some minor evolution can however still be observed at $P=0.01$. 
To gain further simulation speed we combined a linked cell neighbour 
search with a Verlet neighbour list\cite{22}. In the late stages of evolution, when particles hardly 
move, this leads to a large increase in simulation speed, particularly for systems of large 
particle size difference.

\begin{figure}
\includegraphics[width=2.4in,angle=90]{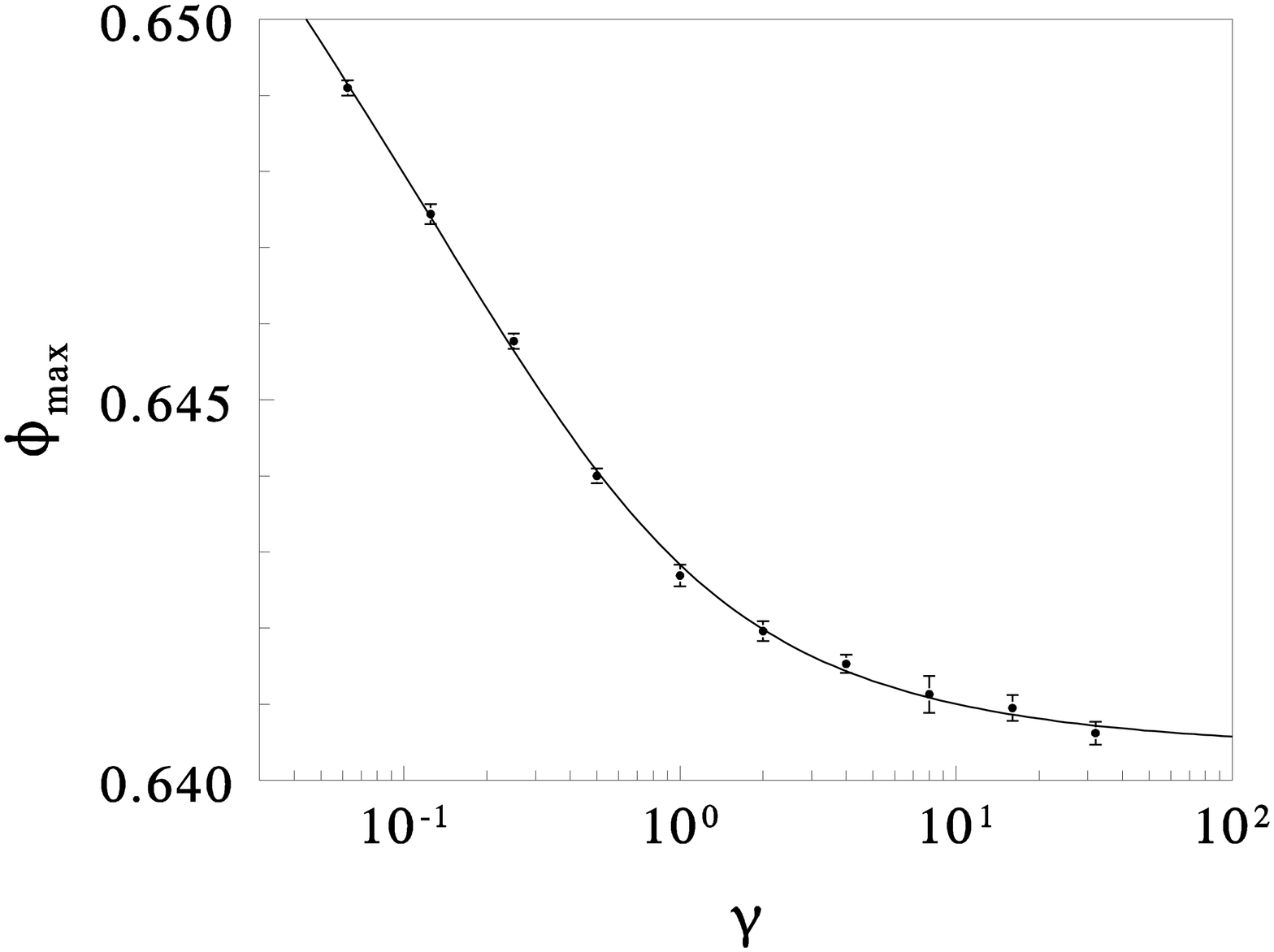}
\includegraphics[width=2.4in,angle=90]{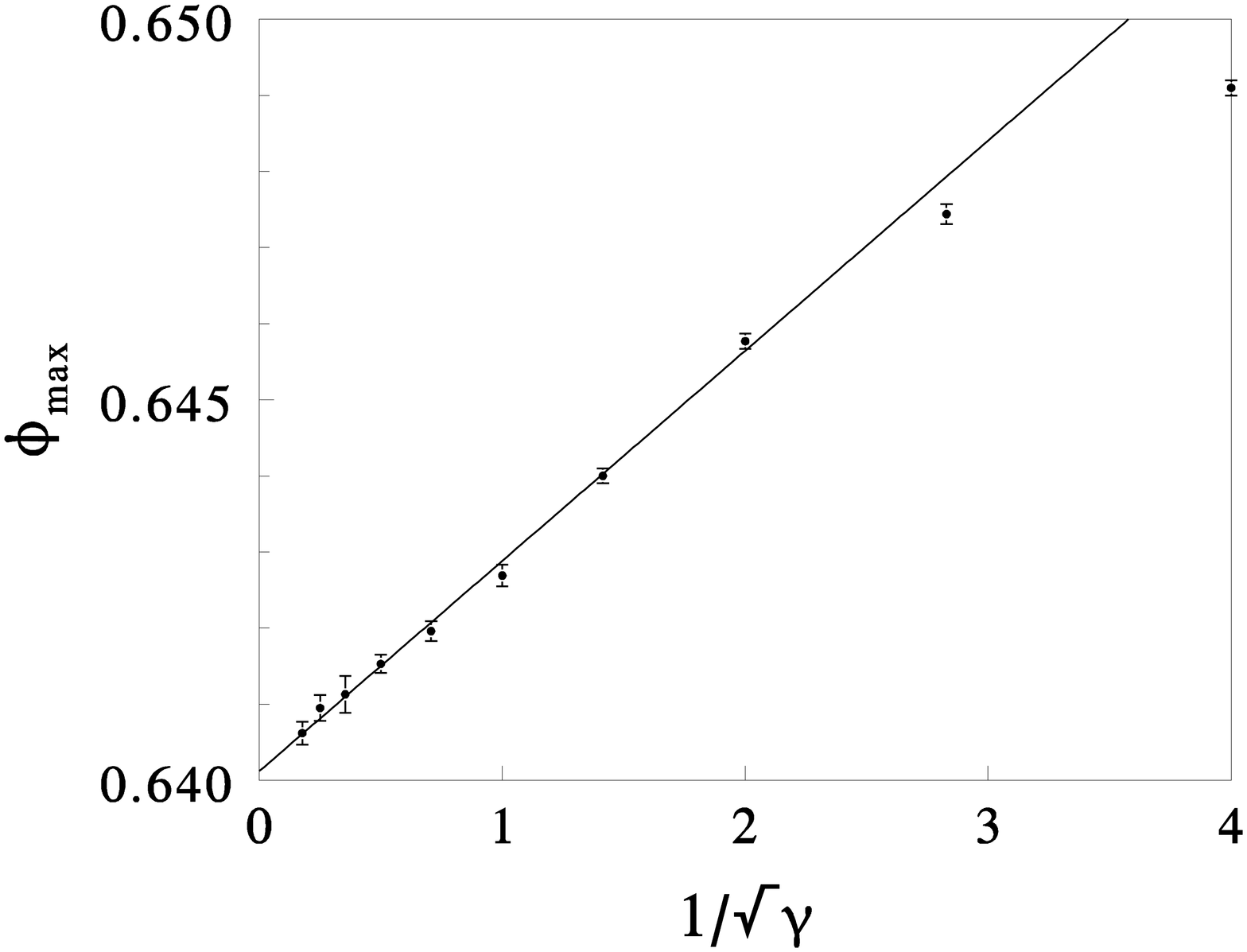}
\caption{\label{fig3}
a (top): Close packing density as function of the friction factor $\gamma$. Each data point is 
an average over 10 independent runs; b (bottom) same data, plotted to $1/\sqrt{\gamma}$. 
}
\end{figure}

Even though the close packing density is now well-defined, it still depends on friction, or the 
cooling rate, see Fig. \ref{fig3}. In fact this is the very source of the previously found dependence
of the close packing density on particle size and mass.
To study this relation, monodisperse systems of 6000 particles 
were used. All particles have diameter 1 
and mass 1, and interact with a repulsive force with $E = 10^{3}$. We insert 
the particles in a box of size $16.83$ (or $\phi=0.66$), pre-equilibrate for $5\times 10^{4}$ 
to $2\times 10^{5}$ time steps of $\delta t = 0.01$ 
until the pressure has fully equilibrated. Then we run a constant pressure 
ensemble, steering the pressure to $P = 0.01$, until the volume fraction 
is stable over four decimal places ($1.5\times 10^{5}$ time steps). All results are 
averaged over 10 runs.  Polycrystalline domains only start to occur 
for $\gamma<0.03$; all systems shown here are isotropic. Over about a decade we find a 
$\log(\gamma)$ dependence for $\gamma\rightarrow 0$ (see Fig. \ref{fig3}a), and over two 
decades we find a power law decay $\phi \approx 0.64 + 0.0028/\sqrt{\gamma}$
for $\gamma>1/4$ (see Fig. \ref{fig3}b). Therefore we have to make a 
choice for the friction factor, and only refer to the packing fraction at that value of $\gamma$. 
Our default value used in the next section is $\gamma =1$. 

For comparison we also evolved a system from its initial conformation to equilibrium using a 
steepest descent method in an $(N, V, T)$ ensemble,
which should compare well with the CG algorithm\cite{16}. The final pressure coincides with our 
result at $\gamma \rightarrow\infty$, which demonstrates that CG simulates the high viscosity limit.
For truly hard spheres the modulus 
diverges, hence the ratio $t_{el}/t_{d} = 2\pi\gamma(D/Em)^{1/2}\rightarrow 0$. 
To simulate this with particles of finite modulus, low values 
of $\gamma$ would be preferred. This implies that the present method is closer to the physical case 
than the CG algorithm to generate hard spheres conformations, unless largely inelastic collisions
are pertinent.

\section{Results}
\begin{figure}
\includegraphics[width=2.5in,angle=90]{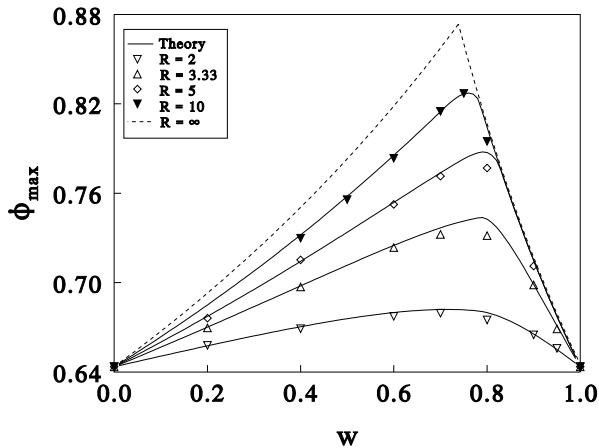}
\caption{\label{fig4}
Maximum packing fraction for bidisperse spheres of different size. 
$R$ is the size ratio, and 
$w = \phi_{\rm large}/(\phi_{\rm large}+\phi_{\rm small})$ is the relative 
volume fraction of the 
large spheres. Symbols are simulation results, and solid lines are theoretical 
predictions, based on 20000 rods. The dashed curves give the upper limit for 
infinite size ratio, Eq. (\ref{eq3}).
}
\end{figure}

The simple approximation for ${\cal F}$ described in section II is now compared with 
the results from the sphere packing simulation method of section III. Consider first bidisperse 
spheres, as studied for example by Clarke and Wiley\cite{23} and by 
Yerazunis et al.\cite{24}, where the larger spheres have $R$ 
times the radius of the smaller, so that 
$P_{3D}(D) \propto R^{3}(1-w) \delta(D-1/R) + w \delta(D-1)$. The 
simulation results for binary mixtures are shown in Table \ref{tab1}, and
in Fig. \ref{fig4}, together with the theoretical 
prediction. The big particles have diameter $D_{1} =$ 1, and the small 
particle diameters are $D_{2} =$ 0.5, 0.3, 0.2 and 0.1. The dash-dot curve 
gives the exact upper limit of the volume 
fraction, Eq. (\ref{eq3}). For diameter $D_{2}\ge 0.2$ we used $6000$ 
particles; for $D_{2} = 0.1$ we used up to $N = 49950$ particles (at $w = 0.8$) 
to prevent finite size effects. All runs were evolved over a 
minimum of 150 000 time steps, and convergence of the volume fraction was 
checked by extending the evolution of selected systems to 450 000 steps. 
All volume fraction results shown in Fig. \ref{fig4} are stable up to four 
decimal places.

\begin{table}
\caption{\label{tab1}Simulation results for the maximum packing fraction of
bidisperse sphers, with diameters $D_{1}$ and $D_{2}$. The mass fraction
present in the large spheres is given by 
$w = \phi_{\rm large}/(\phi_{\rm large}+\phi_{\rm small})$.
}
\begin{tabular}{lcccc}
\hline \hline
$w$ & $D_{2}/D_{1}=0.5$ & $D_{2}/D_{1}=0.3$ & $D_{2}/D_{1}=0.2$ & $D_{2}/D_{1}=0.1$ \\
\hline
0 &
0.6435 &
0.6435 &
0.6435 &
0.6435 \\
0.2 &
0.6579 &
0.6695 &
0.6761 &
\ \\
0.4 &
0.6690 &
0.6971 &
0.7152 &
0.7298 \\
0.5 &
\ &
\ &
\ &
0.7557 \\
0.6 &
0.6774 &
0.7236 &
0.7525 &
0.7835 \\
0.7 &
0.6795 &
0.7324 &
0.7714 &
0.8150 \\
0.75 &
\ &
\ &
\ &
0.8270 \\
0.8 &
0.6749 &
0.7315 &
0.7769 &
0.7948 \\
0.9 &
0.6650 &
0.6985 &
0.7111 &
\ \\
0.95 &
0.6558 &
0.6690 &
\ &
\ \\
1 &
0.6435 &
0.6435 &
0.6435 &
0.6435 \\
\hline \hline
\end{tabular}
\end{table}

A bidisperse system, with moderate to large size ratios, has two distinct 
regimes (and a non-trivial crossover between them): When the proportion 
of large spheres is low, they are isolated from one another, like holes in a 
Swiss cheese; while the small spheres form a 
close-packed phase (the `cheese') between them. On the other hand, when 
the proportion of large spheres is high, these form a close-packed structure, 
leaving the small spheres to 
`rattle around' in the gaps between them. Recent theories of the close packed state of 
bidisperse spheres\cite{14,25} do not address the `rattler' regime adequately. 
In contrast, our theory 
captures both regimes (exactly, in the limit of infinite size ratio), and also the analogous 
regimes which are produced for larger numbers of size classes, such as tridiperse spheres.

\begin{figure}
\includegraphics[width=2.7in]{figure7a.eps}
\includegraphics[width=2.4in,angle=90]{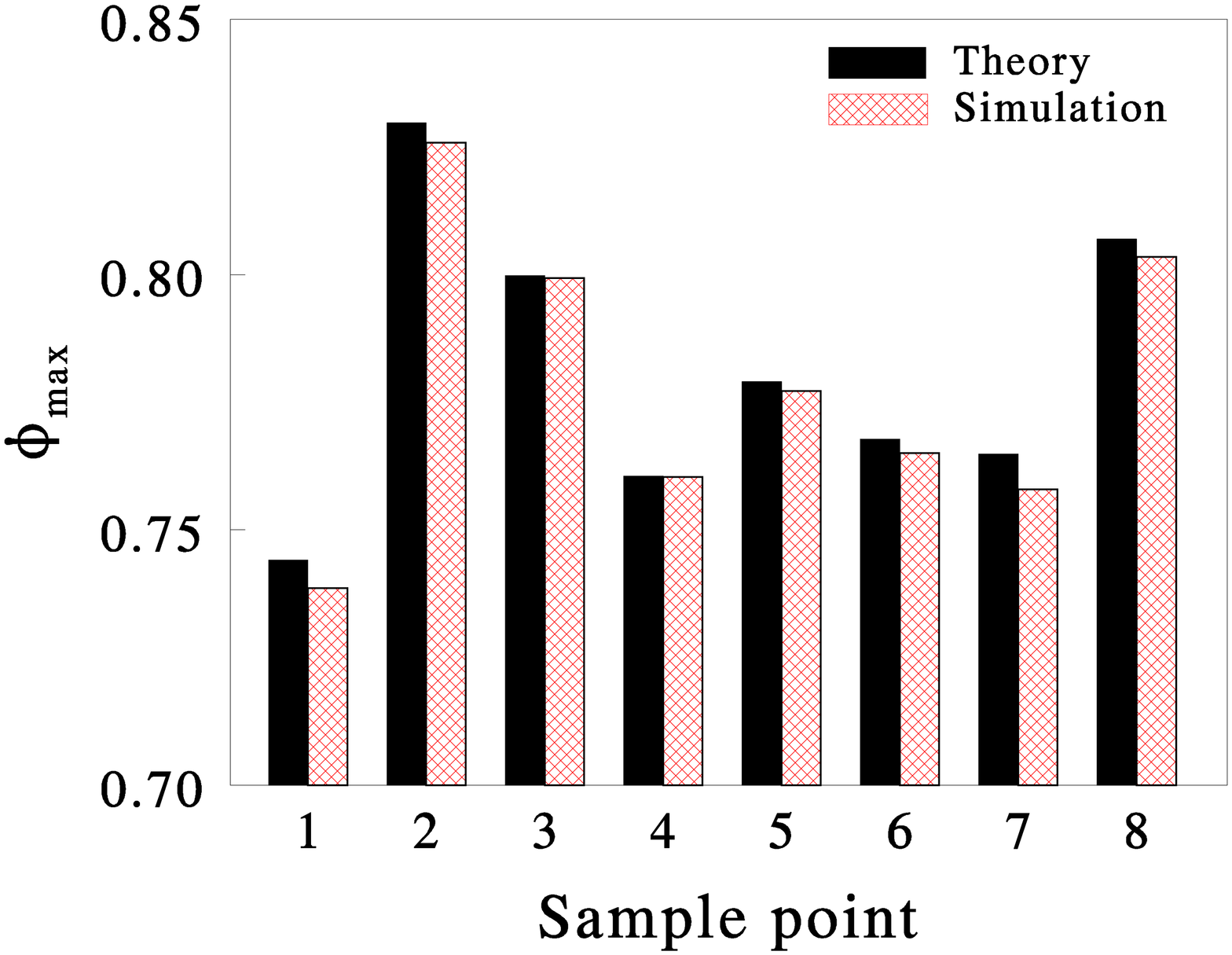}
\caption{\label{fig5}
Maximum packing fraction for a tridisperse distribution of spheres 
with size ratios 1:3:9. The composition diagram (a, top) is based on weight 
fractions; contour lines connect points of equal volume fraction, with bold lines at 
volume fractions of 0.65, 0.7, 0.75 and 0.8. The compositions used in the 
simulations are marked by the circles; b (bottom) shows the comparison 
between theory and simulation per sample point.
}
\end{figure}
          
Next, we consider a tridisperse distribution. In particular we consider the case where 
the sphere diameters are in the ratio 1:3:9. Fig. \ref{fig5} shows the theoretical prediction compared 
with simulation results for selected points. The number of particles chosen varies from 
6000 to 13400. Particular care was taken at large weight fractions of big particles, where 
these may form a stress bearing network. A finite size study showed that in that case 
(specifically sample point 1) the number of large particles in the system needs to be above 
175 for reliable results. For sample point 1 we used 209 big particles from a total of 13300. 
Again, the theory compares very well with the simulation results. The differences may well be
attributed to remaining (minor) finite size effects.
 
\begin{figure}
\includegraphics[width=2.5in,angle=90]{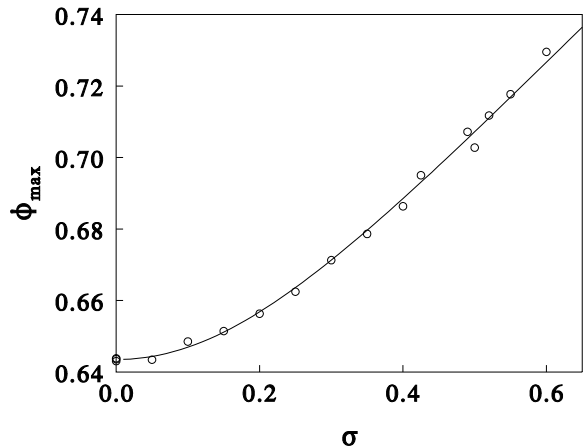}
\caption{\label{fig6}
Maximum packing fraction for a log-normal distribution. The spread 
in log(radius) is denoted by $\sigma$. Simulation results are given by 
symbols; the solid 
line is the theoretical prediction, based on 20000 rods.
}
\end{figure}
     
Finally, we consider a log-normal distribution, which is defined as 
$P_{3D}(D) \propto \exp(- [\ln(D/D_{0})]^{2}/2\sigma^{2})/D$. Thus, 
from Eq. (\ref{eq2}), we find the rod distribution in the theory as 
$P_{1D}(L) \propto L\ {\rm erfc}[\ln(L/D_{0})/\sigma\sqrt{2}]$. In these 
simulations we set an upper diameter to the particles $D = 1$, 
and choose the mean diameter such that less than 0.1\% of the particles 
exceed this size. We 
used 6000 particles that were evolved over 50 000 to 200 000 steps, until the volume 
fraction had converged up to 4 decimal places. Fig. \ref{fig6} shows the comparison between theory 
and simulation data. The system at $\sigma = 0.6$ is shown in Fig. \ref{fig1}. 
Again, the theory compares 
very well with the simulation results.

\section{Conclusions}
Concluding, we have introduced a theory for the close packing density of hard 
spheres of arbitrary size distribution, based on a mapping onto a 
one-dimensional problem. 
To test the theory we simulated the dense random packing of (soft) elastic spheres 
with hydrodynamic friction in 3D in the limit $T\rightarrow 0$, which approaches the hard sphere
system. For the distributions studied we obtain 
excellent agreement between theory and simulation. The theory reproduces the 
infinite size ratio limit for bidisperse spheres. Hence we expect 
that this approximation will prove useful for more general size distributions. 
The simple structure of the approximation may also be amenable to further 
analysis, and open up new avenues for analytical approximations. 

Application of this theory to other space dimensions than 3D is straightforward. However, a comparison
between theory and simulations showed that, although the theory is qualitatively correct in 2D, 
it does not reproduce the simulations as accurately as in 3D. This may be related to the 
mean-field character of the theory, i.e. the explicit spatial correlation is lost in the theory. 
We therefore speculate that the theory will be accurate in 3D and in higher dimensions.

The simulations show a weak dependence of the dense random packing on fluid 
viscosity, if the friction force is of hydrodynamic origin. In general the dense random packing 
density also depends on particle size, mass and elastic modulus.
For particles of diameter $D$, mass $m$ and elastic modulus $E$, suspended in a 
liquid of viscosity $\eta$, we infer that the dense random packing density should be a function of 
the dimensionless group $Q=\eta^2D/(Em)$. For systems of the same $Q$ value but different size, mass and
viscosity we find excellent scaling of the pressure as function of time. Therefore we conclude that
the size dependence of dense random packing is a kinetic effect that
disappears when $m\propto D$. Although such scaling is artificial, it leads to a well-defined
dense random packing, which is prudent to test theories that are based only on geometrical considerations.

\end{document}